\documentclass[12pt]{article}
\usepackage{graphicx}
\usepackage{amssymb}
\usepackage{epstopdf}
\def\cala{{\mathcal A}}
\def\calm{{\mathcal M}}
\def\calt{{\mathcal T}}
\def\call{{\mathcal L}}

\def\pref{{\Omega_{D-3}\over 16\pi G}}
\def\8pig{8\pi G}

\begin{document}

\begin{titlepage}
\vfill
\begin{flushright}
\end{flushright}

\vfill
\begin{center}
\baselineskip=16pt
{\Large\bf The Thermodynamics of}
\vskip 0.3cm
\vskip 0.3cm
{\Large\bf Kaluza-Klein Black Hole/Bubble Chains}
\vskip 1.0cm
{\large {\sl }}
\vskip 10.mm
{\bf David Kastor\footnote{\texttt{kastor@physics.umass.edu}}, 
Sourya Ray\footnote{\texttt{sourya@physics.umass.edu}} 
and Jennie Traschen\footnote{\texttt{traschen@physics.umass.edu}}} \\
\vskip 1cm
{

       Department of Physics\\
       University of Massachusetts\\
       Amherst, MA 01003\\
}
\vspace{6pt}
\end{center}
\vskip 0.5in
\par
\begin{center}
{\bf Abstract}
 \end{center}
\begin{quote}
A Killing bubble is a minimal surface that arises as the fixed surface of a spacelike Killing field.
We compute the bubble contributions to the Smarr relations and the mass and tension first laws for spacetimes containing both black holes and Killing bubbles.  The resulting  relations display an interesting interchange symmetry between the properties of black hole horizons  and those of KK bubbles.  This interchange symmetry reflects the underlying relation between static bubbles and black holes under double analytic continuation of the time and Kaluza-Klein directions. 
The thermodynamics  of
bubbles involve a geometrical quantity that we call the bubble surface gravity, which we
show has several properties in common with the black hole surface gravity.
\vfill
\vskip 2.mm
\end{quote}
\end{titlepage}

\section{Introduction}

Kaluza-Klein (KK) bubbles are fascinating objects.  The original KK bubble  presented by Witten \cite{Witten:1981gj}, which describes the non-perutrbative decay of the KK vacuum, was highly dynamical.   However, many static KK bubble spacetimes are also known, and these too display interesting physical properties.   For example, it was shown in \cite{Elvang:2002br} that two arbitrarily large
black holes can be held apart by a KK bubble.  A large group of spacetimes, containing both black hole horizons and KK bubbles, have been studied in reference \cite{Elvang:2004iz}.  These spacetimes are all static and translation invariant in the Kaluza-Klein direction, and fall within the class of generalized Weyl solutions constructed in reference \cite{Emparan:2001wk}.  They display an interesting interchange symmetry between black holes and KK bubbles that arises  because static black holes and KK bubbles are related through double analytic continuation of the time and KK directions.  Under this analytic continuation, the ADM mass $\calm$ of the spacetime is interchanged with the product $\calt\call$ of the ADM tension and the length of the compact direction measured at infinity.

Given the mechanical interplay between black holes and bubbles, it is natural to ask whether  KK bubbles  also contribute  to the laws of black hole thermodynamics?  In particular, we would like to know whether these laws display the interchange symmetry described above between black holes and bubbles?
We will see that the answer to both these questions is `yes'.

It should be noted that, as discussed in \cite{Elvang:2004iz},  the black hole/bubble interchange symmetry is broken by requiring that the bubbles be smooth.  If the period $\call$ of the Kaluza-Klein direction is fixed, then the parameters associated with the KK bubbles cannot be tuned freely.  Since there is no analogous requirement for the black holes, the smoothness condition breaks the symmetry between them.    Since we are particularly interested in the interchange symmetry, we will not impose the smoothness condition and will allow for conical singularities at the KK bubbles.  
Physically, we can think of a conical bubble as being a smooth bubble wrapped 
by a $p=-\rho$ brane.  A KK bubble is a minimal surface, and therefore the brane equation of motion will be satisfied for such a configuration.   The stress-energy of an idealized wrapped brane creates the conical singularity at the bubble.  A similar construction was used in the context of branes wrapping  Euclidean black hole horizons in reference \cite{Dowker:1991qe}.

In this paper we will derive  Smarr relations for  spacetimes 
containing both black holes and bubbles.  There are two Smarr relations, one corresponding to the time translation Killing vector and a second one corresponding to the spatial translation Killing vector in the KK direction.  These were derived in the absence of KK bubbles in references   \cite{Harmark:2003dg} and \cite{Kastor:2007wr} respectively.
We will also compute the bubble contributions to the first laws satisfied by perturbations between black hole/bubble spacetimes.  In this case there are also two laws to consider, the ordinary first law for variations in the mass $\delta\calm$ and the first law for variations in the tension $\delta\calt$, which was derived in the absence of bubbles in  references \cite{Traschen:2001pb,Townsend:2001rg}.

We collect our results here in the introduction.
The  Smarr relation for the time translation Killing vector, the first equation below,
receives no bubble corrections.  Bubbles do contribute to the Smarr relation resulting from the spatial translation Killing vector, which is given in the second equation
\begin{equation}\label{introsmarr}
{1\over\8pig} \kappa_H \cala_H = {D-3\over D-2}\calm - {1\over D-2}\calt \call,  \qquad 
{1\over\8pig} \kappa_B \cala_B\call = {D-3\over D-2}\calt \call - {1\over D-2}\calm .
\end{equation}
Here $\cala_H$ and $\cala  _B$ are the areas of the black hole horizon and KK bubble 
respectively, $\kappa_H$ is surface gravity of the black hole horizon, and $\kappa_B$ is the KK bubble surface gravity, which we will define below.   The two Smarr formulas obviously satisfy the interchange symmetry between bubbles and black holes discussed above, with $\calm$ being swapped with the product $\calt\call$, $\kappa_H$  with $\kappa_B$, and $\cala_H$  with the product $\cala_B\call$.

We find that the mass and tension first laws including bubble contributions are given by
\begin{eqnarray}\label{introthermo}
 \delta\calt &=& {1\over\8pig}\kappa_B\delta\cala_B-{1\over\8pig\call} \cala_H\delta\kappa_H
 \\  \nonumber
\delta\calm &=& {1\over\8pig} \kappa_H\delta\cala_H - {1\over\8pig} \call\cala_B\delta\kappa_B 
+(\calt- {1\over\8pig}\kappa_B\cala_B)\delta\call.
\end{eqnarray}
If we suppose that the length $\call$ is fixed, then these relations are symmetric 
under the same interchange of quantities as the Smarr relations. 
One intriguing feature of these formulas is the intricate interchange between the  intensive quantities 
$\kappa_H$ and $\kappa_B$ and the extensive quantities $\cala_H$ and $\cala_B$.  Before the bubble contributions are added in, $\delta\calm$ depends on the variations of the extensive $\cala_H$, while 
$\delta\calt$ depends on the variation of the intensive parameter $\kappa_H$.  The bubble quantities enter into the mass and tension first laws with the extensive and intensive contributions reversed.

Allowing $\call$ to vary, the mass first law picks up an additional  work term,
with an effective tension that includes a contribution from the bubble.   This term breaks the interchange symmetry, reflecting the distiction that the spacelike KK direction is periodic, with period $\call$, while the Lorentzian time direction is infinite in extent.

The paper proceeds as follows.  In section (\ref{preliminaries}) we establish some basic conventions and review the properties of the uniform black string and static Kaluza-Klein bubble spacetimes. 
 In section (\ref{why}) we show why bubble contributions to the thermodynamic laws should be expected.  
In section (\ref{bsg}) we discuss necessary aspects of the geometry of KK bubbles, including the notion of Killing bubbles and the definition of the bubble surface gravity.  
We show that the surface gravity is constant over the bubble and that a smooth Killing bubble is a minimal submanifold.  This last result justifies our picture of a conical bubble as a smooth bubble wrapped by a brane.
In section (\ref{smarrsection}) we study the bubble contributions 
to the Smarr relations.  In section (\ref{firstlawsection}), we compute the bubble contributions to the mass and tension first laws.
In section (\ref{example}) we check our results in a simple example containing both a KK bubble and a black hole.

\section{Simple black hole and bubble spacetimes}\label{preliminaries}

We consider spacetimes that at infinity are asymptotic to $M^{D-1}\times S^1$ and assume that the compact KK direction has coordinate $z$ identified according to $z\equiv z+\call$.  Such spacetimes are characterized at infinity by their ADM mass and tension\footnote{For simplicity, we assume that the ADM angular momenta and momentum in the compact direction vanish.}, which 
 are given in terms of the asymptotic forms of the metric coefficients
\begin{equation}
g_{tt} \simeq -1 +c_t / r ^{D-4}  ,\qquad  g_{zz}  \simeq  1 +c_z /r ^{D-4}
\end{equation}
by the formulas \cite{Harmark:2003dg}
\begin{equation}\label{massandtension}
{\cal M} = {\Omega_{D-3}\call\over 16\pi G} ( (D-3) c_t -c_z ),\qquad
  {\cal T}  = {\Omega_{D-3}\over 16\pi G} (  c_t -(D-3)c_z ),
\end{equation}

It is helpful to orient ourselves by considering two simple vacuum solutions, the uniform black string and the static Kaluza-Klein bubble.   Some of the basic properties of these two spacetimes are as follows.
The uniform black string is given by the $D-1$ dimensional Schwarzschild solution crossed with a flat line in the $z$ direction, 
\begin{equation}\label{ubs}
ds^2 = -f(r)dt^2 + dz^2 +{dr^2\over f(r)} +r^2d\Omega_{D-3}^2, \qquad f(r)=1-(c/r)^{D-4}
\end{equation}
The uniform black string has a Killing horizon at $r=c$.
Its mass and tension, together with the horizon area and surface gravity are given by
\begin{equation}\label{ubsinfo}
{\cal M} =(D-3) {\Omega_{D-3}\call\over 16\pi G}  c^{D-4},\qquad
{\cal T}  = {\Omega_{D-3}\over 16\pi G}c^{D-4},\qquad
\cala_H = \Omega_{D-3}\call c^{D-3}, \qquad \kappa_H= {(D-4)\over 2c}
\end{equation}

The static KK bubble is related to the uniform black string by  double analytic continuation of the time and compact coordinates.
It is given by the Euclidean $D-1$ dimensional Schwarzschild solution crossed with a flat time direction, 
\begin{equation}\label{kkb}
ds^2 = -dt^2 + f(r)dz^2 +{dr^2\over f(r)} +r^2d\Omega_{D-3}^2, \qquad f(r)=1-(c/r)^{D-4}
\end{equation}
The compact dimension pinches off at the KK bubble $r=c$, which is a 
codimension $2$ surface.  In the terminology of reference  \cite{Gibbons:1979xm} the KK bubble is a bolt.  Following \cite{Gibbons:1979xm}, we will associate a surface gravity $\kappa_B$ with KK bubbles, which is defined in equation (\ref{bolt}) below.
For the static KK bubble, the mass and tension together with the bubble area and surface gravity are 
\begin{equation}\label{kkbinfo}
\calm = \pref \call c^{D-4},\qquad \calt=(D-3)\pref c^{D-4},\qquad 
\cala_B= \Omega_{D-3} c^{D-3},\qquad \kappa_B= {D-4\over 2c}
\end{equation}
Note that in going from the uniform black string to the static KK bubble, the quantities $\calm$ and $\calt\call$ are interchanged.

For general values of the parameters $c$ and $\call$ the static KK bubble has a conical singularity at $r=c$.  Smoothness of the KK bubble requires a particular relation between $c$ and $\call$, as follows. Although we will not require smoothness of Kaluza-Klein bubbles, we will want to keep track of the smoothness condition.
Setting $r= c +(D-4)y^2/(4c)$ and focusing on the region $y\ll 1$ in (\ref{kkb}) gives
\begin{equation}
ds^2 \simeq  \kappa_B^2 \, y^2 dz^2 +dy^2+(c+{(D-4)y^2 \over 4c})^2d\Omega_{D-3}^2-dt^2
\end{equation}
with $\kappa_B$ as in equation (\ref{kkbinfo}).
The deficit angle is given by 
\begin{equation}\label{deficit}
\psi = 2\pi -\kappa_B\call.
\end{equation}
Hence, the smoothness condition is $\psi=0$ or equivalently  $\kappa_B\call/2\pi=1$.

\section{Why are bubble contributions to black hole thermodynamics necessary?}\label{why}

In this section we show that the first law for black holes in Kaluza-Klein theory needs to be modified for spacetimes including KK bubbles.  The idea will be to show that starting from the first law, without any additional KK bubble contributions, we can derive a Smarr relation that is not satisfied by the basic static KK bubble spacetime  (\ref{kkb}). The first law without bubble contributions \cite{Townsend:2001rg,Harmark:2003eg,Kastor:2006ti} is given by 
\begin{equation}\label{oldmassfirstlaw}
\delta\calm = {1\over\8pig} \kappa_H\delta\cala_H+\calt\delta\call
\end{equation}
For classes of spacetimes having certain types of symmetries, the first law can be used to derive two independent Smarr relations via scaling arguments, as follows.

First consider the Smarr relation for static spacetimes.  Given a static vacuum solution of general relativity, there always exists a one-parameter family of such solutions related to one another by changes in  the overall length scale (see {\it e.g.} reference \cite{Chowdhury:2006qn}).
The physical parameters of these spacetimes scale according to their dimensionalities.  For a KK black hole spacetime, we have a family of solutions in which the  parameters associated with the black hole scale as
$\calm = \lambda^{D-3}\bar\calm$, 
$\cala_H=\lambda^{D-2}\bar\cala_H$ and
$\call=\lambda\bar\call$.
The spacetime may also include a KK bubble, but the variation of its properties under scaling do not enter into equation  (\ref{oldmassfirstlaw}).
Requiring that the first law  (\ref{oldmassfirstlaw}) holds for variations in the overall scale $\lambda$ leads to the Smarr relation 
\begin{equation}\label{smarr1}
{1\over\8pig} \kappa_H \cala_H = {D-3\over D-2}\calm - {1\over D-2}\calt \call
\end{equation}

A second, independent Smarr relation can be derived for spacetimes that are both static and translation invariant in the compact $z$ coordinate. 
In this case, given one such solution there exists a family of solutions related simply by rescaling the length of the  compact direction \cite{Kastor:2007wr}.  If we scale this length according to $\call=\rho\bar\call$, then the physical parameters entering equation (\ref{oldmassfirstlaw}) scale according to
$\calm = \rho\bar\calm$, $\cala_H=\rho\bar\cala_H$.  Requiring that the first law be satisfied for variations in $\rho$, we get the Smarr formula
\begin{equation}\label{smarr2}
\calm = {1\over\8pig} \kappa_H\cala_H + \calt\call
\end{equation}

The uniform black string (\ref{ubs}) and the static KK bubble (\ref{kkb}) are both static and translation invariant in the compact direction, and therefore should each satisfy both Smarr relations (\ref{smarr1}) and (\ref{smarr2}).  It is straightforward to check that this is indeed true for the  uniform black string. However, the KK bubble satisfies  (\ref{smarr1}), but not  (\ref{smarr2}).  We therefore conclude that the first law (\ref{oldmassfirstlaw}), which was the starting point for the derivations of both (\ref{smarr1}) and (\ref{smarr2}), does not hold in the presence of KK bubbles.

We can get a clue as to  the missing element  in the second Smarr relation by making one further manipulation.  The Smarr relations (\ref{smarr1}) and (\ref{smarr2}) can be combined so as to eliminate the horizon area, giving
\begin{equation}\label{smarr3}
0 = {D-3\over D-2} \calt \call- {1\over D-2}\calm .
\end{equation}
The overall factor of $1/D-2$ is included in equation (\ref{smarr3}) is in order to highlight its similarity to equation (\ref{smarr1}).  The right hand sides of the two equations are related by interchanging $\calm$ with $\calt\call$.  Recall that these same two quantities are interchanged in going between the uniform black string and the static KK bubble.  However, the left hand sides of equations (\ref{smarr1}) and (\ref{smarr3}) are not related by any interchange, and it seems reasonable to conjecture that the bubble area and surface gravity will appear on the left hand side of  the corrected version of (\ref{smarr3}) to give an overall duality between the two formulas.

\section{Geometry of Killing bubbles} \label{bsg}

In order to evaluate the bubble contributions to the Smarr formulas and to the mass and tension first laws, we need to study the geometry in a neighborhood of a bubble.  We will focus on the geometry of a Killing bubble, which we take to be the fixed surface of a spacelike Killing field 
$Z^a=(\partial/\partial z)^a$, where $Z^a$  translates around the compact Kaluza-Klein circle.    

We will start by assuming that the spacetime geometry  in a neighborhood of the bubble is smooth and that we can introduce coordinates $x^a$ in this neighborhood such that the bubble is the co-dimension $2$ submanifold $x^I=0$ with $I=1,2$.  The $1$-forms $dx^I$ are then normal to the bubble at $x^I=0$.  The spacetime metric can be written  in the form 
\begin{equation}\label{genmetric}
ds^2 = \gamma_{IJ}dx^Idx^J + \lambda_{\alpha\beta}(dx^\alpha + \rho^\alpha{}_Idx^I)
(dx^\beta + \rho^\beta{}_Jdx^J)
\end{equation}
where Greek indices run over the range $0,3,\dots,D-1$.
We will also use an orthonormal frame given by
\begin{equation}\label{genbasis}
e^{\hat I}= E^{\hat I}{}_J dx^J, \qquad e^{\hat\alpha}= 
E^{\hat\alpha}{}_\beta(dx^\beta + \rho^\beta{}_I dx^I)
\end{equation}
where $\gamma_{IJ}=\delta_{\hat K\hat L}E^{\hat K}{}_I E^{\hat L}{}_J$ and 
$\lambda_{\alpha\beta}= \eta_{\hat\rho\hat\sigma} E^{\hat\rho}{}_\alpha E^{\hat\sigma}{}_\beta$.

It is important to note that  the derivative of the Killing vector $Z^a$ vanishes when projected onto the bubble.  To see this, let $V^a$ be a vector tangent to the bubble.
Since the vector $Z^a$ vanishes identically on the bubble, it follows  that $V^a \partial _a Z_b $      and hence that $V^a \nabla _a Z_b  = V^a \nabla _{[a} Z_{b]}=0$.
Therefore, the derivative of $Z^a$ at the bubble can be written as
\begin{equation}\label{bolt}
\nabla _{[a} Z_{b]}  = 
\kappa_B (e^{\hat 1}_a e^{\hat 2}_b -\hat{ e}^{\hat 1}_b e^{\hat 2}_a )  .  
\end{equation}
We will refer to equation (\ref{bolt}) as the bolt equation using the terminology of reference 
\cite{Gibbons:1979xm}.  We define the bubble surface gravity to be the coefficient 
$\kappa_B$ in the bolt equation.
At this point it appears that $\kappa_B$   could be a non-trivial function of the coordinates 
$x^\alpha$ along the bubble.  However, one can show that $\kappa_B$ is constant, via the following steps.

Contracting  the bolt equation  with itself gives the useful expression
\begin{equation}\label{kappadefthree}
\kappa ^2 _B ={1\over 2}  (\nabla _{[a} Z_{b]} )\nabla ^{[a} Z^{b]} ,
\end{equation}
where the right  hand side is understood to be  evaluated on the bubble at $x^I=0$.
Equation (\ref{kappadefthree}) is quite similar to a standard expression for the black hole surface gravity $\kappa_H$, which is obtained by flipping the overall sign on the right hand side and replacing $Z^a$ with the horizon generating Killing field.  Now, take the derivative of the right hand side of 
(\ref{kappadefthree}) in a direction tangent to the bubble.
Making use of the facts  that 
$\nabla_a\nabla_b Z_c = - R_{bcad}Z^d$ for any Killing vector and that $Z^a$ vanishes on the bubble, we see that the derivative of the right hand side vanishes.  Therefore $\kappa_B$ is constant on the bubble.
Given its similarities with the horizon surface gravity $\kappa_H$, it is natural to refer to $\kappa_B$  as the bubble surface gravity.

Equation (\ref{kappadefthree}) for $\kappa_B$ will be used in the computation of the boundary term in the Smarr formula.  Determining the bubble contributions to the mass and tension first laws will require yet another equivalent expression for $\kappa_B$
\begin{equation}\label{kappaagain}
\kappa_B^2 = \lim_{x^I\rightarrow 0}\left ( {(\nabla_a Z\cdot Z)(\nabla^a Z\cdot Z)\over 4 Z\cdot Z}\right ),
\end{equation}
where the limit is necessary because both the numerator and denominator vanish on the bubble.  
In order to derive equation (\ref{kappaagain}) one notes that the bolt equation implies that near the bubble,
\begin{equation}
{1\over 2}\nabla_a(Z\cdot Z) = \kappa_B Z^b
(e^{\hat 1}_a e^{\hat 2}_b -\hat{ e}^{\hat 1}_b e^{\hat 2}_a ) +\dots
\end{equation}
where the terms neglected vanish more rapidly near $x^I=0$.  With this ingredient, construct the right hand side of (\ref{kappaagain}) expressing the $e^{\hat I}$ in terms of the metric coefficients.  Taking the limit $x^I\rightarrow 0$ then gives equation (\ref{kappaagain}).

Let $\rho_a=\nabla_a(Z\cdot Z)$ and $\hat \rho_a = \rho_a/\sqrt{\rho\cdot\rho}$.  
The $1$-form $\hat\rho_a$ is then orthogonal to $Z^a$ and normal to surfaces of constant $Z\cdot Z$.
We can then further process (\ref{kappaagain}) into the statement
\begin{equation}\label{kappayetagain}
\kappa_B = \lim_{x^I\rightarrow 0} \hat\rho^a\, \nabla_a(\sqrt{Z\cdot Z}).
\end{equation}
This will be useful in proving the first laws.

We can also use equation (\ref{bolt}) to show that a Killing bubble has vanishing extrinsic curvature.
The bubble is a codimension $2$ submanifold, and therefore has two extrinsic curvature tensors, given by
\begin{equation}
K_{ab}^{\hat I} = h_a{}^c\nabla_c e^{\hat I}_b,\qquad I=1,2
\end{equation}
where $h_{ab}$ is the metric on the bubble, $h_{ab}= g_{ab} - \delta_{\hat I\hat J}e^{\hat I}_a e^{\hat J}_b$.  Now consider taking a derivative of equation (\ref{bolt}) in a direction tangent to the bubble.   The derivative of the left hand side will be zero, for the same reasons given above, while the derivative of the right hand side can be expressed in terms of the extrinsic curvature tensors.   On the bubble, we then have
\begin{eqnarray}
0 = h_c{}^d\nabla_d\nabla_aZ_b  &=& \kappa_B\, h^c{}^d\nabla_d
(e^{\hat 1}_a e^{\hat 2}_b -e^{\hat 1}_b e^{\hat 2}_a )  \\ \nonumber
& = & \kappa_B ( K_{ca}^{\hat 1} e^{\hat 2}_b + K_{cb}^{\hat 2} e^{\hat 1}_a
-K_{cb}^{\hat 1} e^{\hat 2}_a - K_{ca}^{\hat 2} e^{\hat 1}_b).
\end{eqnarray}
Contracting with $e^{\hat 1 a}$ and $e^{\hat 2a}$ respectively then implies that $K_{cb}^{\hat 2}=K_{cb}^{\hat 1}=0$.  Smooth Killing bubbles are therefore minimal submanifolds.  The equations of motion for the Nambu-Goto area action are solved by a static brane wrapping a minimal submanifold.  Therefore, as discussed in the introduction, Killing bubbles with conical singularities can be thought of as smooth Killing bubbles wrapped by branes.

\subsection{Normal bubbles}\label{conical}

As we will see in section (\ref{massfirstlawsection}) below, smooth bubbles do not contribute to the mass first law.  However, bubbles with conical singularities do give nontrivial contributions.  Including these contributions will reveal the symmetry between the mass and tension first laws under the interchange of black hole and bubbles.

Consider the $2$-dimensional submanifold with coordinates $(\rho,z)$ where $\rho = \sqrt{Z\cdot Z}$, and the bubble is at $\rho=0$.  Here we will assume that the spacetime metric near $\rho=0$ is block diagonal, $ds^2 \simeq d\sigma^2 + B_{\alpha\beta}dx^\alpha dx^\beta$ with
\begin{equation}\label{2dmetric}
d\sigma^2 \simeq d\rho^2 +\kappa_B^2\rho^2 dz^2
\end{equation}
and $0<\kappa_B\le 2\pi/\call$.  That is, the two dimensional submanifold is either smooth at the bubble with $\kappa_B = 2\pi/\call$, or it has a deficit angle $\psi = 2\pi - \kappa_B\call$.  We will refere to these as `normal bubbles' because of the block diagonal structure of the metric.   It can be checked that the black hole/bubble example considered in section (\ref{example}) is a normal bubble, as is the basic static KK bubble in equation (\ref{kkb}).

A priori, $\kappa_B$ is some parameter.  However, substituting into the definition  (\ref{kappaagain}) one can verify that $\kappa_B$ in (\ref{2dmetric}) is indeed the bubble surface gravity.   Hence, we see that for normal Killing bubbles, the value of $\kappa_B$ is fixed in terms of $\call$ and $\psi$,
\begin{equation}
\kappa_B = {2\pi\over\call}(1 - {\psi\over 2\pi}).
\end{equation}
One might wonder how generic this relation is?  That is to say, are bubbles usually normal?  This turns out to require a more detailed analysis of the near-bubble geometry, which we defer to future work.

\section{Smarr formulas via Komar integral relations}\label{smarrsection}

The  Smarr formulas, including KK bubble contributions,  are derived starting from  the geometric Komar integral relations. 
As discussed in the introduction, we will see that with the bubble contribution included, the two Smarr relations (\ref{smarr1}) and equation (\ref{zsmarr}) below are dual to one another under the double analytic continuation interchanging uniform black strings and KK bubbles.

A  Killing vector $k^a$ satisfies  
 $-\nabla _c \nabla ^c k^a = R^a {}_b k^b = 8\pi (T^a {}_b - g^a {}_b T/(D-2) )$, where the first equality
 is a geometric identity, and the second holds for solutions to the Einstein equations. Let $V$ be a 
 co-dimension one hypersurface  with unit normal $n_a$, and boundaries  $\partial V_i$. Integrating over $V$ gives the Komar integral relation    $\sum_i I_{\partial V_i}= 8\pi\int _V dv k^b n_a
  (T^a {}_b - g^a {}_b T/(D-2) )$, where
\begin{equation}\label{komarrelation}
I_S= -{1\over 16\pi G}\int_S dS_{ab} \nabla^ak^b.
\end{equation}
For the most part, we will be focused on vacuum spacetimes, for which  $\sum_i I_{\partial V_i}=0$.
However, we will find an application for the formula with stress energy in the case of  bubbles with deficit angle.

We first compute the bubble corrections to
the Smarr formula (\ref{smarr3}), which was derived in the absence of KK bubbles in reference 
\cite{Kastor:2007wr}.  Consider a spacetime that is static and $z$-translation invariant.  Take the Killing vector in  (\ref{komarrelation}) to be the $z$-translation Killing vector $Z^a$, and integrate over the boundaries of a surface of constant $z$. 
The additional assumption of staticity is necessary in order to be able to cancel
the contributions from the initial and the final constant time boundaries. 
The remaining boundaries of the region of integration are  at infinity, at the black hole horizon,
and at the bubble.  The boundary integral at the horizon vanishes.  This comes about in the following way.  Let  $l^a$ be the Killing generator of the horizon.  The normal form $dS_{ab}$ to the black hole horizon within a constant $z$ slice will be proportional to $Z_{[a}l_{b]}$ and the integral (\ref{komarrelation}) at the black hole horizon will therefore include the quantity
\begin{equation}
Z_{[a}l_{b]} \nabla^a Z^b = k^a k^b\nabla_{[a} l_{b]} = 0,
\end{equation}
where we have used the fact that the two Killing vectors commute.

Including a KK bubble introduces a new boundary in the interior of a constant $z$ slice.
 By definition the Killing vector 
$\partial/\partial z$ vanishes at a Kaluza-Klein bubble.  Surfaces of constant $z$ come together at the bubble, so for any given surface of constant $z$ we must introduce an additional boundary at a bubble.  Assume that there is only one bubble present.  For more than one bubble, the result is simply the sum of contributions from each individual bubble\footnote{We have implicitly made a similar assumption, that there is a single connected black hole horizon.  For more than one black hole, the result is summed over contributions from the individual black hole horizons.}.

Now consider the boundary term at the bubble.  The bolt equation (\ref{bolt}) implies that $\nabla_a Z_b$ is normal to the bubble in both its indices.  Since the bubble has codimension $2$, we see that $\nabla_a Z_b$ 
must be proportional to the volume form $dS_{ab}$.  The correct normalization gives
\begin{equation}
dS_{ab}= {\sqrt{2}\, \nabla_a Z_b\over \sqrt{(\nabla^c Z^d)(\nabla_c Z_d)} }
\end{equation}
Making use of  formula (\ref{kappadefthree}) for the bubble surface gravity and using the fact that 
$\kappa_B$ is constant on the bubble,  we then arrive at the result
\begin{equation}
I_B=-{1\over 8\pi G}\kappa_B\cala_B
\end{equation}
where $\cala_B$ is the area of the bubble.
  The right hand side of (\ref{smarr3}) represents the contribution from the boundary at infinity, which is unchanged in this case.  
As in reference  \cite{Kastor:2007wr} we have
\begin{eqnarray}
I_{\infty} &=& -  {\Omega_{D-3}\over 16\pi G} (D-4)c_z\\
&=&  {1\over D-2} ((D-3)\calt - \calm/\call ) .
\end{eqnarray}
The equation $I_\infty + I_B +I_H =0$ then gives the Smarr formula 
\begin{equation}\label{zsmarr}
{1\over\8pig} \kappa_B \cala_B\call = {D-3\over D-2}\calt \call - {1\over D-2}\calm
\end{equation}
Equation (\ref{zsmarr}) has been multiplied by an overall factor of $\call$ to facilitate comparison with equation (\ref{smarr1}).    

Next we show that equation (\ref{smarr1}) does not receive any new contributions from KK bubbles.  One might expect this to be the case, since (\ref{smarr1}) is satisfied by the static KK bubble, and indeed is true even if the static bubble is not smooth, {\it i.e.} with the parameters $c$ and $\call$ chosen independently.  
In the absence of Kaluza-Klein bubbles, the Smarr formula (\ref{smarr1}) was obtained for static spacetimes by taking $k^a$ to be the time translation Killing vector $l^a$ on a constant time hypersurface, with boundaries at infinity and the black hole horizon  \cite{Harmark:2003dg} .
If the bubble is smooth, then since $l^a$ is timelike on the bubble (unlike a black hole horizon),
there is no need for  a boundary at the bubble, and so clearly there is no contribution. If there is a deficit
angle at the bubble, then we need to check whether that introduces a contribution.   We do this in the following way.

As discussed above, 
the metric with deficit angle can be considered to be a simple model for the metric outside
a brane that wraps the bubble.
Let us replace the conical bubble metric,
by a smooth metric with the stress energy of the wrapped brane.  
There is now no boundary at the bubble, but there is
 a contribution to the  Komar integral relation from the volume integral over  stress energy  terms.
The brane is described by an area action $S_{brane} =\rho \int \sqrt{-B}$, where the integral
is over the worldvolume of the brane, and $B_{ab}$ is the induced metric on the worldvolume. The stress energy tensor
is given by $T_{ab} =-\rho B_{ab}$ and has  trace $T= -(D-2)\rho$.
If we now evaluate on the brane the combination $T^a {}_b - B^a {}_b\, T/(D-2) $ that contributes to the Komar relation, we find that it vanishes identically, and hence there is no stress-energy
contribution from the conical bubble. Note that the cancellation to give zero only works for a co-dimension $2$ brane, as we have here.
We comment that, in light of the results of \cite{Geroch:1987qn}, one generally needs to be cautious about using models with stress-energy concentrated on submanifolds of co-dimension greater than one in general relativity.   However, in the present case, the high degree of symmetry together with the 
$p=-\rho$ equation of state, combine to give sensible results.

Having found the bubble contribution to the left hand side of (\ref{zsmarr}) and established that equation (\ref{smarr1}) does not receive any bubble contribution, we see that there is now a duality, or interchange symmetry, between the two Smarr formulas, reflecting the interchange of black holes with bubbles under double analytic continuation of the $z$ and $t$ coordinates.

\section{Mass and tension first laws for KK bubbles}\label{firstlawsection}

In this section, we will derive the mass and tension first laws for KK bubble spacetimes using the Hamiltonian perturbation theory techniques of reference \cite{Sudarsky:1992ty}\cite{Traschen:2001pb}. We assume that the spacetime can be foliated by a family of hypersurfaces $V$ specified by the unit normal $w^a$, and that the surfaces can be described as level surfaces of a coordinate $w$. We also assume that the spacetime metric $\bar{g}_{ab}$ is a solution to the vacuum Einstein equations with a Killing vector $k^{a}$, which we decompose into normal and tangential components to the hypersurfaces $V$, according to $k^a=Fw^a+\beta^a$.   The mass and tension first laws follow from making different choices of the Killing vector $k^a$ and the normal field $w^a$.

Now we consider that the metric
$g_{ab} = \bar{g}_{ab}+\delta{g}_{ab}$ is the linear approximation to another solution to the vacuum Einstein equations. Denote the Hamiltonian data for the background metric
by $(\bar{s}_{ab}$, $\bar{\pi}_{ab})$, the corresponding perturbations  by $h_{ab} = \delta \bar{s}_{ab}$ and $p^{ab} = \delta \bar{\pi}^{ab}$, and the linearized Hamiltonian and momentum constraints by
$\delta H$ and $\delta H_a$. The following statement then holds as a consequence of the constraint Einstein's equations and  Killing's equation in the background metric \cite{Traschen:1984bp}
\begin{equation}\label{divgauss}
F\delta H+\beta^a \delta H_a=-\bar{D}_aB^a
\end{equation}
Here $\bar{D}_a$ is the background covariant derivative operator compatible with $\bar{s} _{ab}$.
 The vector $B^a$ is given by 
\begin{equation}\label{boundaryterm}
B^a=F(\bar{D}^ah-\bar{D}_bh^{ab})-h\bar{D}^aF+h^{ab}\bar{D}_bF+{1 \over \sqrt{|\bar{s}|}}\beta^b(\bar{\pi}^{cd}h_{cd}\bar{s}^{a}_{\ b}-2\bar{\pi}^{ac}h_{bc}-2p^a_{\ b})
\end{equation}
Indices are raised and lowered with the background metric $\bar{s}_{ab}$. For solutions to the vacuum Einstein equations, we know that $\delta H=\delta H_a=0$. Therefore, 
equation (\ref{divgauss}) becomes a Gauss' law type statement $\bar{D}_aB^a=0$. Integrating over the hypersurface $V$ and applying Stokes theorem gives
\begin{equation}\label{boundaryintegral}
\sum_i \left( \int_{\partial V_I}da_cB^c\right) =0
\end{equation}
where $\partial V_i$ are the disjoint components of the boundary of the hypersurface $V$. For black hole spacetimes, for example, the boundary of the hypersurface $V$ typically has two components, one on the event horizon and one at infinity. 
 
Smooth bubbles such as (\ref{kkb}) do not introduce interior boundaries to the spacetime. However, for particular foliations of the spacetime, the Hamiltonian evolution may not be well defined at the bubble. 
In order to make use of the Hamiltonian perturbation theory formalism, it is then necessary to introduce as in reference \cite{Hawking:1995fd} an inner boundary surrounding the bubble.   There can then be non-trivial contributions from this boundary to equation (\ref{boundaryintegral}).  Inner boundary contributions can also arise from perturbations that have conical singularities at the bubble.  
  
\subsection{Bubble contributions to the tension first law}

 The tension first law requires that the background spacetime have a spatial translation symmetry along the compact $z$ direction.  As above, we take the $z$ coordinate have period $\call$. 
 We choose $V$ to be a hypersurface of constant $z$, and suppose
 that the spacetime contains a bubble and possibly also a  black hole horizon.  Constant $z$ hypersurfaces meet at the bubble and consequently the Hamiltonian flow is not defined there. As in \cite{Hawking:1995fd}  it is then necessary to introduce an inner boundary surrounding the intersection of $V$ with the bubble. Let $\partial V_{B}$ , $\partial V_H$, and $\partial V_{\infty}$ denote the  boundaries of the hypersurface $V$ at the bubble, at the horizon, and  at infinity respectively.   
 
Integration over $V$ includes the time direction.  We will take this into account by considering, as in reference 
 \cite{Traschen:2001pb},  only background spacetimes $\bar g_{ab}$ and perturbations $\delta g_{ab}$ that are static.  We then integrate over a portion of $V$ between some arbitrary initial and final times.  The assumption of staticity implies that the boundary terms on the initial and final surfaces of $V$ cancel.
Equation (\ref{boundaryintegral}) then implies that
\begin{equation}\label{bubbleboundary}
I_B +I_H +I_{\infty}=0
\end{equation}
where
$I_B=\int_{\partial V_B}da_cB^c$,
and similarly for $I_H$ and $I_{\infty}$.

The boundary term at infinity was computed in reference \cite{Traschen:2001pb} and is given by
 \begin{equation}
 I_{\infty}=-(16\pi G)\, \delta{\cal T}\, \Delta t.
 \end{equation} 
 Here $\delta\calt$ is the perturbation to the ADM tension and 
$\Delta t$ is the time interval between the initial and final time slices in the region of integration.
 
Now consider the boundary term at the bubble.  Near the bubble we choose $V$ such that its normal is proportional to the Killing vector $Z^a=(\partial / \partial z)^a$. The normalization factor $F$ is then given by $F=\sqrt{Z\cdot Z}$.   This factor vanishes on the bubble, and therefore the only non-vanishing terms in (\ref{boundaryterm}) are those proportional to the derivative of  $F$. Let $\hat{\rho}^c$ be the unit  normal to the  bubble within $V$.  From equation (\ref{kappayetagain}) we have that
 $\hat{\rho}^c\, \nabla_cF=\kappa_B$.  
The boundary term at the bubble is then given by
 \begin{eqnarray}
 I_B&=&-\int_{\partial V_B}da\hat{\rho}_c(-h\bar{D}^aF+h^{ab}\bar{D}_bF) \nonumber \\
 &=& 2\kappa_B\, \delta {\cal A}_B\, \Delta t
 \end{eqnarray}
 where ${\cal A}_B$ is the area of the bubble at constant time,
 and we have used the result that $\kappa_B$ is constant on the bubble. The minus sign
 in the first line is because the outward normal to $\partial V$ points in the $-\hat{\rho}_c$ direction.
 If there is a black hole in the spacetime, then there will be an additional contribution from the horizon boundary, which was found in reference \cite {Traschen:2001pb} to be
 \begin{equation}
 I_{H}=-{2\cala_H \delta \kappa_H \over \call}\Delta t.
 \end{equation}
 Putting all the boundary terms together as in equation (\ref{bubbleboundary}) and dividing out the common factor of  the time interval $\Delta t$, we arrive at the tension first law for black hole/bubble spacetimes
 \begin{equation}\label{tensionfirstlaw}
 \delta\calt = {1\over\8pig}\kappa_B\delta\cala_B-{1\over\8pig\call} \cala_H\delta\kappa_H.
 \end{equation}
Note that the bubble term in the tension first law resembles the black hole horizon term in the mass first law.

 \subsection{Bubble contributions to the mass first law}\label{massfirstlawsection}
 
 We now turn to the derivation of the mass first law. Let 
the background be a static bubble spacetime, possibly also including a black hole and take the Killing vector in the Hamiltonian perturbation theory construction to be the time translation Killing field
 $l^a=(\partial / \partial t)^a$.  The hypersurface $V$ is taken to be spacelike and to approach 
 a constant $t$ slice at infinity.   A smooth bubble does not introduce any new boundary into the interior of $V$, and therefore smooth perturbations to smooth bubbles do not lead to bubble contributions to the mass first law.  Nontrivial bubble contributions come only from perturbations that either introduce, or change the size of the deficit angle\footnote{For conical bubble backgrounds, one can show that perturbations that do not change the strength of the conical singularity give vanishing contributions to the mass first law.} in the plane orthogonal to the bubble as described in section (\ref{conical}).  In this case, one must introduce a boundary surrounding the bubble, which excises the conical singularity.  In the remainder of this section, we will restrict our attention to normal bubbles as defined in section (\ref{conical}).

Accordingly, we consider perturbations that change the parameter $\kappa_B$ in equation (\ref{2dmetric}).  Near the bubble, these have the form
\begin{equation}
h_{zz}\simeq 2\kappa_B\, \delta\kappa_B\, \rho^2
\end{equation}
Introduce an inner boundary surrounding the bubble, evaluate the integral (\ref{boundaryintegral}) and then take the limit as the boundary shrinks to the bubble. A smooth field will not contribute in the limit.
 Evaluating the terms in (\ref{boundaryterm}),  the only non-zero contribution comes from
 $\bar{D}_bh^{\rho b}=-2\delta \kappa_B/(\kappa_B\rho)$.
The integral on the boundary surrounding the bubble then gives
 \begin{eqnarray}
 I_B(\delta\kappa_B)&=&\int_{\partial V_B}da\hat{\rho}_c F\bar{D}_bh^{cb} \nonumber\\ 
  &=& - 2\cala_B\call\delta \kappa_B
 \end{eqnarray}
 where we have used the fact that the product $Fda$ gives the volume element on the bubble, which then integrates to give
 the area of the bubble.
 
There is a second independent contribution to the mass first law that comes from varying the range of the $z$ coordinate.   As discussed in \cite{Kastor:2006ti}, the contribution of variations $\delta\call$  to the first law can be handled via a coordinate transformation, so that $\delta \call$ appears in the metric perturbation, rather than in the range of coordinate $z$. Following this procedure, yields the metric perturbation $h_{z z}=(2\delta \call / \call )\bar g_{zz}$ with $z$ identified according to $z\equiv z +\call$.
Computing the boundary integral at the bubble then gives
\begin{equation}
I_B(\delta \call) =    - 2\cala_B\kappa_B \delta\call
\end{equation}

We also need to include the contribution from a possible black hole horizon, which has the standard form $I_H=2\kappa_H\delta \cala_H$.
Finally, there is the  boundary integral at infinity which was shown in \cite{Kastor:2006ti} to be $I_\infty=-16\pi G(\delta \calm-\calt \delta \call)$.
Collecting all these boundary terms together, we obtain the mass first law for black hole/bubble spacetimes
\begin{eqnarray}\label{massfirstlaw}
\delta\calm &=& {1\over\8pig} \kappa_H\delta\cala_H - {1\over\8pig} \call\cala_B\delta\kappa_B 
+(\calt- {1\over\8pig}\kappa_B\cala_B)\delta\call \\
&=& {1\over\8pig} \kappa_H\delta\cala_H +\calt\delta\call -{1\over 8\pi G}\cala_B\delta(\kappa_B\call)
\end{eqnarray}
We emphasize that $\delta\kappa_B=0$ for smooth perturbations.  The $\delta\kappa_B$ contribution
comes only from changes in the deficit angle, as defined in section (\ref{conical}).
From the second form of the equation, we see that the bubble contribution to the mass first law is indeed proportional to the variation in the deficit angle $\psi$ in the plane normal to the bubble.  This is clear since $\kappa_B\call = 2\pi -\psi$.  As discussed in the introduction, with the bubble contributions included the mass and tension first laws, equations 
(\ref{massfirstlaw}) and (\ref{tensionfirstlaw}), now display the interchange symmetry between bubbles and black holes.

\section{Example: 
a black hole/bubble chain}\label{example}

We have now derived the bubble contributions to the mass and tension first laws (\ref{massfirstlaw}) and (\ref{tensionfirstlaw}), as well as the Smarr formulas (\ref{smarr1}) and (\ref{zsmarr}).  It is straightforward to show that all these results hold in the static KK bubble spacetime (\ref{kkb}).  In this section, we will check that they hold in a more intricate example that includes both a KK bubble and a black hole horizon.

A large class of $5$ dimensional black hole bubble chains is presented and discussed in reference  \cite{Elvang:2004iz}.  In these spacetimes, black hole horizons alternate with Kaluza-Klein bubbles.  We will focus on the simplest one, which contains just one of each.  The metric for this spacetime is given by

\begin{equation}
ds^2=-e^{2U_1}dt^2+e^{2U_2}dz^2+e^{2U_3}d\psi^2+e^{2\nu}(dR^2+d\phi^2)
\end{equation}
where
\begin{eqnarray}
e^{2U_1}&=&\frac{R_1-\zeta_1}{R_2-\zeta_2}, \hspace{25pt} e^{2U_2}=\frac{R_2-\zeta_2}{R_3-\zeta_3}, \hspace{25pt}
e^{2U_3}=(R_1+\zeta_1)(R_3-\zeta_3), \nonumber\\
e^{2\nu}&=&{1\over2^{3/2}R_1R_2R_3}\sqrt{Y_{13}Y_{12}Y_{23}}\frac{R_3-\zeta_3}{R_1-\zeta_1}, \nonumber \\ 
\zeta_i&=&\phi-a_i, \hspace{25pt}
R_i=\sqrt{R^2+\zeta_i^2}, \hspace{25pt}
Y_{ij}=R_iR_j+\zeta_i\zeta_j+R^2, 
\end{eqnarray}
where  in our notation $z$ is the coordinate on the circle $S^1$. The geometry of the spacetime is such that on the $R=0$ surface the black hole horizon extends from $\phi=a_1$ to $\phi=a_2$ whereas the bubble extends from $\phi=a_2$ to $\phi=a_3$. The rod structure, as discussed in \cite{Elvang:2004iz}, for this configuration is given in  figure (\ref{compare C}) below.
\begin{figure}[ht]
\centerline{\includegraphics[width=.6\textwidth]{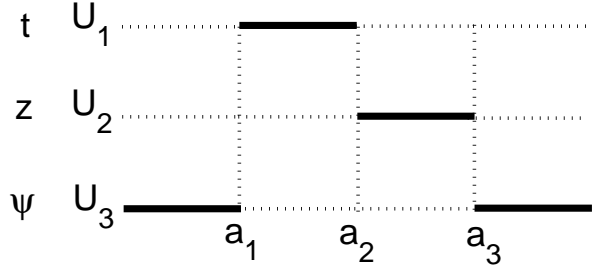}}
\caption{Rod structure for the black hole/bubble spacetime.  The finite rods in the $t$ and $z$ directions correspond to the black hole and KK bubble respectively, while the semi-infinite rods in the $\psi$ direction correspond to axes of rotation.}
\label{compare C}
\end{figure}
%


To verify the Smarr relations and the first laws we need to obtain the expressions for the suface gravity and the area of the bubble in terms of the paramaters of the metric. First we note that the 3 parameters  $a_1$, $a_2$ and $a_3$ can be reduced to 2 parameters $b_1$ and $b_2$ using translational invariance in the $\phi$-direction. In other words, one can perform the coordinate transformation $\phi \rightarrow \phi+(a_2+a_3)/2$ suitable for analyzing the near-bubble geometry. Then the new parameters are related to the old ones by the relations $b_1=a_1-(a_2+a_3)/2$ and $b_2=(a_3-a_2)/2$.
Next we apply a coordinate transformation from the Weyl coordinates $(R,\phi)$ to the polar coordinates $(r, \theta)$ given by
\begin{eqnarray}
\label{coordtransf1}
R=r \sin{\theta} \sqrt{1-{2b_2\over r}}, \hspace{35pt} \phi=(r-b_2)\cos{\theta}
\end{eqnarray}
The metric now takes the form
\begin{equation}
\label{polarform}
ds^2=-e^{2U_1}dt^2+e^{2U_2}dz^2+e^{2U_3}d\psi^2+e^{2\nu}(\sigma_{\theta\theta}d\theta^2+\sigma_{rr}dr^2)
\end{equation}
where
\begin{eqnarray}
\sigma_{\theta\theta}=r(r-2b_2)+b_2^2\sin^2{\theta}, \hspace{25pt}\sigma_{rr}=\sin^2{\theta}\frac{(r-b_2)^2}{r(r-2b_2)}+\cos^2{\theta}
\end{eqnarray}
In these coordinates, the location of the bubble is at $r=2b_2$. In the limit $r \rightarrow 2b_2$, the various metric coefficients upto the leading order are
\begin{eqnarray}
 g_{tt}&\simeq&-\frac{b_2(1+\cos{\theta})}{b_2\cos{\theta}-b_1}, \hspace{10pt} g_{zz}\simeq{r-2b_2 \over 2b_2}, \hspace{10pt}
 g_{\psi\psi}\simeq4b_2(1-\cos{\theta})(b_2\cos{\theta}-b_1) \nonumber\\ g_{rr}&\simeq&-\frac{b_2-b_1}{r-2b_2}, \hspace{10pt}
 g_{\theta\theta}\simeq-2b_2(b_2-b_1)
\end{eqnarray}
This has the form of the metric for a single static bubble in equation (\ref{kkb}) with $r\rightarrow c$.
Vanishing of $g_{zz}$ on the bubble shows that the Killing vector $Z^a=({\partial \over \partial z})^a$ vanishes on the bubble surface. Using the expression (\ref{kappaagain}), the surface gravity of the bubble is given by
\begin{equation}\label{kappab}
\kappa_B^2=\lim_{\rho \rightarrow 2b_2} {1\over 4} {g^{rr} \over g_{zz} } (\partial _r g_{zz})^2
= {1\over 8b_2(b_2-b_1)}.
\end{equation}

The area $\cala_B$ of the bubble is found to be
\begin{eqnarray}\label{Ab}
{\cal A}_B&=&\lim_{\rho\rightarrow 2b_2}\int d\psi\ d\theta\ (e^{2U_1}e^{2U_3}e^{2\nu}\sigma_{\theta\theta})^{1/2} \nonumber\\
&=&8\pi b_2\sqrt{2b_2(b_2-b_1)}
\end{eqnarray}
Now, with the above expressions for the surface gravity $\kappa_B$ and the area $\cala _B$ of the bubble we can check the second Smarr relation (\ref{zsmarr}) for this particular configuration of a black hole on a bubble in $D=5$.
Using the expressions (\ref{massandtension}) for the ADM mass and tension in terms of the asymptotic forms of the metric coefficients in $D=5$ dimensions, the right hand side of the Smarr relation (\ref{zsmarr}) becomes $-c_{z}\call/ 4G$. In case of sequence of bubbles and black holes written in the generalized Weyl form, the parameter $-c_{z}$ is given by the sum of the lengths of the rods sourcing the bubbles, which in the present parametrization is just $2b_2$ and the parameter $c_t$ is given by the sum of the lengths of the rods sourcing the event horizons, which in the present parametrization is $-(b_1+b_2)$. The left hand side of (\ref{zsmarr}) can be computed using the expressions (\ref{kappab}) and (\ref{Ab}) to obtain $b_2\call/2G$. 
 This checks the second Smarr relation.
 
 Next we check the first laws for this solution, including  conical bubbles.
  We do this by varying all the physical quantities with respect to the three parameters $b_1,b_2$ and $\call$. We will need the values of $\kappa _H$ and ${\cal A}_H$ in terms of the $b_i$.
  To analyze the near-horizon geometry  we choose a  different parametrization than the previous one.  Let $c_1=(a_2-a_1)/2$ and $c_2=a_3-(a_1+a_2)/2$
and perform a similar coordinate transformation as (\ref{coordtransf1})
\begin{eqnarray}
\label{coordtransf2}
R=r \sin{\theta} \sqrt{1-{2c_1\over r}}, \hspace{35pt} \phi=(r-c_1)\cos{\theta}.
\end{eqnarray}
 The metric again takes the form (\ref{polarform}) with $\sigma_{\theta\theta}$ and $\sigma_{rr}$ given by 
 \begin{eqnarray}
\sigma_{\theta\theta}= r(r-2c_1)+c_1^2\sin^2{\theta}, \hspace{25pt}\sigma_{rr}=\sin^2{\theta}
\frac{(r-c_1)^2}{r (r-2c_1)}+\cos^2{\theta}
\end{eqnarray}
 In this parametrization, the location of the event horizon is at $r=2c_1$. In the limit $r \rightarrow 2c_1$, the various metric coefficients are
\begin{eqnarray}
 g_{tt}&\simeq&-{r-2c_1 \over 2c_1},\hspace{10pt}
 g_{zz}\simeq \frac{c_1(1-\cos{\theta})}{c_2-c_1\cos{\theta}},\hspace{10pt}
 g_{\psi\psi}\simeq 4c_1(1+\cos{\theta})(c_2-c_1\cos{\theta}),\nonumber \\ 
 g_{rr}&\simeq& \frac{c_1+c_2}{r-2c_1},\hspace{10pt}
 g_{\theta\theta}\simeq2c_1(c_1+c_2)
\end{eqnarray}
The area of the event horizon and the horizon surface gravity are then found to be
\begin{eqnarray}
{\cal A}_H&=&\lim_{\rho\rightarrow 2c_1+\epsilon}\int d\phi\ d\psi\ d\theta\ (e^{2U_2}e^{2U_3}e^{2\nu}\sigma_{\theta\theta})^{1/2} 
=8\pi Lc_1\sqrt{2c_1(c_1+c_2)}\nonumber\\
\kappa_H &=&{1\over 2\sqrt{2c_1(c_1+c_2)}}.
\end{eqnarray}
These can be rewritten in terms of the parameters  $b_1$ and $b_2$,with the result
 \begin{equation}
 \label{Akappa}
 \kappa_H={1 \over 2\sqrt{b_1^2-b_2^2}}; \hspace{25pt}
 {\cal A}_H=4\pi \call(-b_1-b_2)^{3/2}(b_2-b_1)^{1/2}.
 \end{equation}
This is in agreement with the expressions given in \cite{Elvang:2004iz}.

The mass first law is
\begin{equation}
\delta M -{\mathcal T}\delta \call={1 \over 8 \pi G}[\kappa_H \delta {\cal A}_H-\call {\cal A}_B\delta \kappa_B-\kappa_B{\cal A}_B \delta \call]
\end{equation}
First we note that the mass and the tension are given in terms of the parameters $b_1 <0$ and $b_2$ as
 \begin{equation}
 M=-{\call b_1 \over 2G};\hspace{25pt}{\mathcal T}={3b_2-b_1 \over 4G}
 \end{equation}
 Thus the left hand side of the relation gives
 \begin{equation}
 \delta M -{\mathcal T}\delta \call=-{\call \over 2G}\delta b_1-{3b_2+b_1 \over 4G}\delta \call
 \end{equation}
 Now, we use the expressions for $\kappa_H, \kappa_B, \cala_H$ and $\cala_B$ from (\ref{kappab}),(\ref{Ab}) and (\ref{Akappa}), to evaluate the right hand side of the mass first law.
 \begin{equation}
 {1 \over 8 \pi G}[\kappa_H \delta {\cal A}_H-\call{\cal A}_B\delta \kappa_B-\kappa_B{\cal A}_B \delta \call]=-{\call \over 2G}\delta b_1-{3b_2+b_1 \over 4G}\delta \call
 \end{equation}
 which verifies the   mass first law for the black hole-bubble pair.

Similarly, the variation in tension is given as
\begin{eqnarray}
\delta \calt={1 \over 4G}(3\delta b_2-\delta b_1).
\end{eqnarray}
Again, using the expressions for $\kappa_H, \kappa_B, \cala_H$ and $\cala_B$ from (\ref{kappab}),(\ref{Ab}) and (\ref{Akappa}) to evaluate the right hand side of the tension first law, we get
\begin{eqnarray}
{1\over\8pig}\kappa_B\delta\cala_B-{1\over\8pig\call} \cala_H\delta\kappa_H={1 \over 4G}(3\delta b_2-\delta b_1)
\end{eqnarray}
thus checking the tension first law.

\section{Conclusions}\label{conclusions}

We have studied the KK bubble contributions to the Smarr relations and mass and tension first laws.  With the bubble contributions included,  the two Smarr relations as well as the two first laws satisfy an interchange symmetry, between black hole and bubble contributions, that follows essentially from the exchange of black holes and bubbles under double analytic continuation. It is interesting that
this symmetry extends to the wrapping of branes in the two cases.  The wrapping of classical
branes around KK bubbles has an analog in the wrapping of quantum vortices around Euclidean
black hole horizons \cite{Dowker:1991qe}.

There are several open questions. The mass first law depends on the perturbation $\delta\kappa_B$ to the bubble surface gravity.
Since we have only defined $\kappa_B$ for Killing  bubbles, this means that
the mass first law is restricted to perturbations that preserve translation invariance.  
To extend the range of validity, one needs a definition of bubbles and their surface gravity
for non-Kiling bubbles. Work in progress indicates that this is possible, and 
involves further understanding near-bubble geometry.  

More fundamentally,
it would be interesting to 
have thermodynamic interpretations for the bubble contributions to the Smarr relations and to the mass and tension first laws.  One intriguing connection between minimal surfaces, recalling that bubbles are minimal surfaces, and thermodynamics comes from recent work in the $AdS/CFT$ context \cite{Ryu:2006bv}.  
It is conjectured that  one-quater the area of a minimal surface (with boundary) in Anti-deSitter 
spacetime, is equal to the entanglement entropy of a conformal field theory defined on the
boundary.  At this point however, the thermodynamic role of $\cala _B$ and $\kappa_B$ is
an open question.

\subsection*{Acknowledgements}
DK and JT  thank Henriette Elvang for helpful conversations.  This work was supported in part by NSF grant PHY-0555304.


\begin{thebibliography}{99}

\bibitem{Witten:1981gj}
  E.~Witten,
  ``Instability Of The Kaluza-Klein Vacuum,''
  Nucl.\ Phys.\  B {\bf 195}, 481 (1982).


\bibitem{Elvang:2002br}
  H.~Elvang and G.~T.~Horowitz,
  ``When black holes meet Kaluza-Klein bubbles,''
  Phys.\ Rev.\  D {\bf 67}, 044015 (2003)
  [arXiv:hep-th/0210303].


\bibitem{Elvang:2004iz}
  H.~Elvang, T.~Harmark and N.~A.~Obers,
  ``Sequences of bubbles and holes: New phases of Kaluza-Klein black holes,''
  JHEP {\bf 0501}, 003 (2005)
  [arXiv:hep-th/0407050].
  
\bibitem{Emparan:2001wk}
  R.~Emparan and H.~S.~Reall,
  ``Generalized Weyl solutions,''
  Phys.\ Rev.\  D {\bf 65}, 084025 (2002)
  [arXiv:hep-th/0110258].


\bibitem{Dowker:1991qe}
  F.~Dowker, R.~Gregory and J.~H.~Traschen,
  ``Euclidean black hole vortices,''
  Phys.\ Rev.\  D {\bf 45}, 2762 (1992)
  [arXiv:hep-th/9112065].


\bibitem{Harmark:2003dg}
  T.~Harmark and N.~A.~Obers,
  ``New phase diagram for black holes and strings on cylinders,''
  Class.\ Quant.\ Grav.\  {\bf 21}, 1709 (2004)
  [arXiv:hep-th/0309116].
  
\bibitem{Kastor:2007wr}
  D.~Kastor, S.~Ray and J.~Traschen,
  ``The First Law for Boosted Kaluza-Klein Black Holes,''
  JHEP {\bf 0706}, 026 (2007)
  [arXiv:0704.0729 [hep-th]].
  
\bibitem{Traschen:2001pb}
  J.~H.~Traschen and D.~Fox,
  ``Tension perturbations of black brane spacetimes,''
  Class.\ Quant.\ Grav.\  {\bf 21}, 289 (2004)
  [arXiv:gr-qc/0103106].
  
\bibitem{Townsend:2001rg}
  P.~K.~Townsend and M.~Zamaklar,
  ``The first law of black brane mechanics,''
  Class.\ Quant.\ Grav.\  {\bf 18}, 5269 (2001)
  [arXiv:hep-th/0107228].

  
\bibitem{Gibbons:1979xm}
  G.~W.~Gibbons and S.~W.~Hawking,
  ``Classification Of Gravitational Instanton Symmetries,''
  Commun.\ Math.\ Phys.\  {\bf 66}, 291 (1979).

    
\bibitem{Harmark:2003eg}
  T.~Harmark and N.~A.~Obers,
  ``Phase structure of black holes and strings on cylinders,''
  Nucl.\ Phys.\  B {\bf 684}, 183 (2004)
  [arXiv:hep-th/0309230].
   
\bibitem{Kastor:2006ti}
  D.~Kastor and J.~Traschen,
  ``Stresses and strains in the first law for Kaluza-Klein black holes,''
  JHEP {\bf 0609}, 022 (2006)
  [arXiv:hep-th/0607051].

\bibitem{Chowdhury:2006qn}
  B.~D.~Chowdhury, S.~Giusto and S.~D.~Mathur,
  ``A microscopic model for the black hole - black string phase transition,''
  Nucl.\ Phys.\  B {\bf 762}, 301 (2007)
  [arXiv:hep-th/0610069].
  
  
\bibitem{Geroch:1987qn}
  R.~Geroch and J.~H.~Traschen,
  ``Strings and Other Distributional Sources in General Relativity,''
  Phys.\ Rev.\  D {\bf 36}, 1017 (1987).

  
\bibitem{Sudarsky:1992ty}
  D.~Sudarsky and R.~M.~Wald,
  ``Extrema of mass, stationarity, and staticity, and solutions to the Einstein
  Yang-Mills equations,''
  Phys.\ Rev.\  D {\bf 46}, 1453 (1992).
  
\bibitem{Traschen:1984bp}
  J.~H.~Traschen,
  ``Constraints On Stress Energy Perturbations In General Relativity,''
  Phys.\ Rev.\  D {\bf 31}, 283 (1985).


\bibitem{Hawking:1995fd}
  S.~W.~Hawking and G.~T.~Horowitz,
  ``The Gravitational Hamiltonian, action, entropy and surface terms,''
  Class.\ Quant.\ Grav.\  {\bf 13}, 1487 (1996)
  [arXiv:gr-qc/9501014].


\bibitem{Ryu:2006bv}
  S.~Ryu and T.~Takayanagi,
  ``Holographic derivation of entanglement entropy from AdS/CFT,''
  Phys.\ Rev.\ Lett.\  {\bf 96}, 181602 (2006)
  [arXiv:hep-th/0603001].



%

%
%
%



%
%
%
%
 

\end{thebibliography}
 \end{document}